\documentclass[aps,prd,preprint,preprintnumbers,superscriptaddress,showpacs]{revtex4-1}
\usepackage{graphicx}

\usepackage{ulem}
\usepackage{color}
\definecolor{My_red}        {cmyk}{0.00,1.00,1.00,0.20}

\newcommand{\bmat}{\left(\begin{array}}
\newcommand{\emat}{\end{array}\right)}
\newcommand{\beq}{\begin{equation}}
\newcommand{\eeq}{\end{equation}}

\def\bwt{\begin{widetext}}
\def\ewt{\end{widetext}}
\def\be{\begin{equation}}
\def\ee{\end{equation}}
\def\bea{\begin{eqnarray}}
\def\eea{\end{eqnarray}}
\def\bean{\begin{eqnarray*}}
\def\eean{\end{eqnarray*}}
\def\bary{\begin{array}}
\def\eary{\end{array}}
\def\bit{\begin{itemize}}
\def\eit{\end{itemize}}

\def\su5u1{SU(5) \times U(1)}
\def\fsu5u1{SU(5) \times U(1)'}
\def\so10{SO(10)}
\def\sq20{SO(10) \times SO(10)}

\def\bwt{\begin{widetext}}
\def\ewt{\end{widetext}}
\def\be{\begin{equation}}
\def\ee{\end{equation}}
\def\bea{\begin{eqnarray}}
\def\eea{\end{eqnarray}}
\def\bean{\begin{eqnarray*}}
\def\eean{\end{eqnarray*}}
\def\bary{\begin{array}}
\def\eary{\end{array}}
\def\bit{\begin{itemize}}
\def\eit{\end{itemize}}

\def\su5u1{SU(5) \times U(1)}
\def\fsu5u1{SU(5) \times U(1)'}
\def\so10{SO(10)}
\def\sq20{SO(10) \times SO(10)}

\usepackage[centertags]{amsmath}
\usepackage{amssymb}

\begin{document}

\title{The Diboson Excesses in Leptophobic $U(1)_{\rm LP}$ Models \\ from String Theories}

\author{Tianjun Li}

\affiliation{State Key Laboratory of Theoretical Physics and 
Kavli Institute for Theoretical Physics China (KITPC),
Institute of Theoretical Physics, Chinese Academy of Sciences, 
Beijing 100190, P. R. China}

\affiliation{
School of Physical Electronics, University of Electronic Science and Technology of China, 
Chengdu 610054, P. R. China 
}

\author{James A. Maxin}

\affiliation{Department of Physics and Engineering Physics,
  The University of Tulsa, Tulsa, OK 74104, USA}

\author{Van E. Mayes}

\affiliation{Department of Physics, University of Houston-Clear Lake,
  Houston, TX 77058, USA}

\author{Dimitri V. Nanopoulos}

\affiliation{George P. and Cynthia W. Mitchell Institute for Fundamental Physics
  and Astronomy, Texas A$\&$M University, College Station, TX 77843, USA}

\affiliation{Astroparticle Physics Group, Houston Advanced Research Center (HARC),
  Mitchell Campus, Woodlands, TX 77381, USA}

\affiliation{Academy of Athens, Division of Natural Sciences, 28 Panepistimiou Avenue,
  Athens 10679, Greece}

\date{\today}

\begin{abstract}

The ATLAS Collaboration has reported excesses in the search for resonant diboson production with decay modes to hadronic final states at a diboson invariant mass around 2 TeV in boosted jets from $WZ$, $W^+W^-ˆ'$, and $ZZ$ channels. Given potential contamination, we investigate
the anomalies in leptophobic $U(1)_{\rm LP}$ models. We show that leptophobic models can be constructed in flipped $SU(5)\times U(1)_X$
models from free fermionic string constructions and Pati-Salam models from D-brane constructions. Additionally, we perform a collider phenomenological analysis to study production cross sections for $pp \to Z' \to jj/t \bar{t}/WW/Zh$ and discover the excess can be interpreted in both the leptophobic flipped $SU(5)\times U(1)_X$ models and intersecting D-branes.

\end{abstract}


\preprint{ACT-08-15}

\maketitle

\section{Introduction}

The ATLAS and CMS Collaborations have completed searches for massive resonances decaying into a pair of weak gauge bosons via jet substructure techniques, {\it i.e.}, the $pp \to V_1 V_2 \to 4j$ ($V_{1,2}=W^\pm$ or $Z$) channels~\cite{Aad:2015owa, Khachatryan:2014hpa, Khachatryan:2014gha}. The  ATLAS analyses consisted of 20.3 ${\rm fb}^{-1}$ of data at 8 TeV LHC beam collision energies, indicating excesses for narrow widths around $2$~TeV in the $WZ$, $WW$, and $ZZ$ channels with local significances of 3.4$\sigma$, 2.6$\sigma$, and
2.9$\sigma$, respectively~\cite{Aad:2015owa}. Furthermore, CMS performed similar searches, though did not distinguish between $W$-
and $Z$-tagged jets, uncovering a $1.4\sigma$ excess near 1.9 TeV~\cite{Khachatryan:2014hpa}. It is intriguing that CMS also reported about 2$\sigma$ and 2.2$\sigma$ excesses near 1.8 TeV and 1.8--1.9 TeV in the dijet resonance channel and the $e\nu b {\bar b}$ channel, respectively, which could be accounted for by a $W' \to W h$ process~\cite{Khachatryan:2015bma, CMS-Preprint}. Though these excesses are not yet statistically significant, consideration is warranted for potential interpretations of these anomalous events as new physics beyond the Standard Model (SM), as evidence mounts for a possible non-trivial explanation. In the intervening time since ATLAS and CMS first reported their findings, these diboson excesses have been extensively studied~\cite{Fukano:2015hga, Hisano:2015gna, Gerosa:2015tea, Cheung:2015nha, Dobrescu:2015qna,Alves:2015mua, Gao:2015irw, Thamm:2015csa, Brehmer:2015cia, Cao:2015lia, Cacciapaglia:2015eea, Abe:2015jra,Allanach:2015hba, Abe:2015uaa, Carmona:2015xaa, Chiang:2015lqa, Cacciapaglia:2015nga, Fukano:2015uga, Sanz:2015zha,Chen:2015xql, Omura:2015nwa, Anchordoqui:2015uea, Chao:2015eea, Bian:2015ota, Kim:2015vba, Lane:2015fza, Faraggi:2015iaa,Low:2015uha, Liew:2015osa, Terazawa:2015bsa, Arnan:2015csa, Niehoff:2015iaa, Goncalves:2015yua, Fichet:2015yia,Petersson:2015rza, Deppisch:2015cua,Aguilar-Saavedra:2015rna,Bian:2015hda,Dev:2015pga,Franzosi:2015zra}.

The ATLAS diboson excess is well fit by resonance peaks around 2~TeV and widths less than about 100~GeV. Narrow resonances such as this might imply new weakly interacting particles, therefore we shall consider the underlying theories to be perturbative in this work. Turning our focus to the ATLAS excess in the $WZ$, $WW$, and $ZZ$ channels, the tagging selections for each mode used in the analysis are rather incomplete, as these channels share about 20\% of the events. It may be difficult to pronounce that a single resonance is responsible for all excesses, although there does remain the possibility that one 2 TeV particle contributes to the excess in only one channel, whereas the additional excesses in the alternate channels are via contaminations. Approaching the analysis from this perspective provides motivation for not attempting to formulate a simultaneous explanation for all excesses, thereby studying only models with a new resonance in one channel. The reference ranges of the production cross-section times the decay branching ratio for the 2 TeV resonances in the $WZ$, $WW$, and $ZZ$ channels are approximately  $4-8$~fb, $3-7$~fb, and $3-9$~fb, respectively.

The goal in this work is to understand the diboson excesses in leptophobic $U(1)_{\rm LP}$ models from string theories. The leptophobic property aids the process of relaxing LHC search constraints on the leptonic decay channel $Z' \to \ell^+ \ell^-$, with string model building allowing for a deeper understanding of the particle physics. Consequently, we shall realize a leptophobic $U(1)_{\rm LP}$ in flipped $SU(5)\times U(1)_X$ models from free fermionic string constructions~\cite{F-SU5, search, goodies, HFM, Lopez:1996ta, Jiang:2006hf} and in Pati-Salam models from D-brane constructions~\cite{CSU, Cvetic:2004ui, Chen:2005mj,Chen:2006gd, Chen:2006ip,Chen:2007px, Chen:2007zu, Maxin:2011ne, Chen:2007ms, Chen:2005aba, Chen:2005mm, Chen:2005cf}. We conclude the study with exploration of the production cross sections for $pp \to Z' \to jj/t \bar{t}/WW/Zh$, demonstrating a plausible interpretation of the ATLAS excess in both leptophobic flipped $SU(5)\times U(1)_X$ models and intersecting D-branes.

\section{The Leptophobic $U(1)_{\rm LP}$ Model from Stringy Flipped $SU(5)\times U(1)_X$ Models }

The convention we adopt here denotes the SM left-handed quark doublets, right-handed up-type quarks, right-handed down-type quarks, left-handed lepton doublets, right-handed charged leptons, and right-handed neutrinos as $Q_i$, $U_i^c$, $D_i^c$, $L_i$, $E_i^c$, and $N_i^c$ respectively. Our analysis shall investigate the leptophobic $U(1)_{\rm LP}$ model from string theory as a viable explanation of the diboson excess.
The leptophobic $U(1)_{\rm LP}$ cannot be realized in $SU(5)$ models due to the fact the matter field representations $\mathbf{10}_i$ contain $\{Q_i,~U_i^c,~E_i^c \}$, while the representations $\mathbf{\overline{5}}_i$ contain $\{D_i^c, ~L_i \}$. Similar results are found for traditional $SO(10)$ and $E_6$ models as well. Of significant note, representations for three families of SM fermions in flipped $SU(5)\times U(1)_X$ models~\cite{F-SU5} are
\bea
F_i={\mathbf{(10, 1)}}= \{Q_i,~D_i^c,~N_i^c \},~
{\bar f}_i={\mathbf{(\bar 5, -3)}}= \{U_i^c, ~L_i \},~
{\bar l}_i={\mathbf{(1, 5)}}= \{E_i^c\},
\label{smfermions}
\eea
where $i=1, 2, 3$. Notice that $F_i$ does not contain the charged leptons, thus the leptons can be charged under the leptophobic $U(1)_{\rm LP}$ gauge symmetry. It is also clear that ${\bar f}_i$ and ${\bar l}_i$ cannot be charged under the leptophobic $U(1)_{\rm LP}$ gauge symmetry.

In this work we consider flipped $SU(5)\times U(1)_X$ models from four-dimensional free fermionic string constructions~\cite{search}, which
possess various favorable properties regarding vacuum energy, string unification, dynamical generation of all mass scales, top-quark mass, and the strong coupling~\cite{goodies}. The complete gauge group has three identifiable pieces $G=G_{\rm obs}\times G_{\rm hidden}\times G_{\rm U(1)}$,
where $G_{\rm obs}=\rm SU(5)\times U(1)_X$, $G_{\rm hidden}=\rm SU(4)\times SO(10)$, and $G_{\rm U(1)}=\rm U_1(1)\times
U_2(1)\times U_3(1)\times U_4(1)\times U_5(1)$. There are 63 massless matter fields present, annotated in detail in Tables~\ref{Table1}, \ref{Table2}, \ref{Table3}, and \ref{Table4}, including their charges under $G_{\rm U(1)}$. In particular, there are five $F$, two ${\bar F}$, three ${\bar f}$, and three ${\bar l}_i$, which according to the original conventions, are denoted as $F_0$, $F_1$, $F_2$, $F_3$, $F_4$, ${\bar F}_4$, ${\bar F}_5$, ${\bar f}_2$, ${\bar f}_3$, ${\bar f}_5$, ${\bar l}_2$, ${\bar l}_3$, and ${\bar l}_5$, respectively~\cite{search}.

Special emphasis is warranted for the property $\rm Tr\,U_{1,2,3,5}\not=0$, whereas $\rm Tr\,U_4=0$. The anomalous symmetries are artifacts of the truncation of the full string spectrum down to the massless sector. The low-energy effective theory is correctly specified by rotating all anomalies into a single anomalous $\rm U_A\propto \sum_{i=1,2,3,5}\, [{\rm Tr}\,U_i]U_i$~\cite{HFM}, then adding a one-loop correction to the D-term corresponding to $\rm U_A$: $\rm D_A\to D_A+\epsilon M^2$, where $M$ is the reduced Planck scale and $\epsilon=g^2{\rm Tr\,U_A}/192\pi^2$~\cite{DSW}.

The mass spectrum of all states in the accompanying Tables can be obtained through a complex procedure by considering trilinear and non-renormalizable contributions to the superpotential, and likewise to the masses and interactions \cite{search}. However, this procedure does not provide a unique outcome since the VEVs of the singlet fields in Table~\ref{Table2} are unknown, though constrained by the anomalous $\rm U_A$ cancellation conditions. The objective here is to generate an electroweak-scale spectrum that is closely comparable to the MSSM.  Relevant studies can be found in Refs.~\cite{search, Lopez:1996ta}. Two scenarios are studied here that possess an anomaly free leptophobic $U(1)_{\rm LP}$ gauge symmetry~\cite{search, Lopez:1996ta}.

\begin{table}[htp]
\caption{The massless matter fields and their transformation
  properties under $G_{\rm U(1)}$ in the observable sector.
  Under $SU(5)\times U(1)_X$ gauge symmetry, these fields transform
as $F=(10,~1)$, $\bar f=(\bar 5,~-3)$, ${\bar l}^c=(1,~5)$,
$h=(5,~-2)$, and $\bar h=(\bar 5,~2)$. Moreover, we present their charges under
$\rm U_A$ and three orthogonal linear combinations of interest ($\rm
U',U'',U'''$).}
\label{Table1}
\smallskip
\begin{center}
\footnotesize
\begin{tabular}{|l|rrrrr|rrrr|}\hline
&$\rm U_1$&$\rm U_2$&$\rm U_3$&$\rm U_4$&$\rm U_5$&$\rm U_A$
&$\rm U'$&$\rm U''$&$\rm U'''$\\
\hline
$F_0$&$-{1\over2}$&0&0&$-{1\over2}$&0&${3\over2}$&$-{1\over2}$&$-{1\over2}$&0\\
$F_1$&$-{1\over2}$&0&0&${1\over2}$&0&${3\over2}$&$-{1\over2}$&$-{1\over2}$&0\\
$F_2$&0&$-{1\over2}$&0&0&0&${1\over2}$&0&${3\over2}$&$-{1\over2}$\\
$F_3$&0&0&${1\over2}$&0&$-{1\over2}$&${3\over2}$&1&$-{1\over2}$&0\\
$F_4$&$-{1\over2}$&0&0&0&0&${3\over2}$&$-{1\over2}$&$-{1\over2}$&0\\
$\bar F_4$&${1\over2}$&0&0&0&0&$-{3\over2}$&${1\over2}$&${1\over2}$&0\\
$\bar F_5$&0&${1\over2}$&0&0&0&$-{1\over2}$&0&$-{3\over2}$&${1\over2}$\\
$\bar f_2,~{\bar l}^c_2$&0&$-{1\over2}$&0&0&0&${1\over2}$&0
&${3\over2}$&$-{1\over2}$\\
$\bar f_3,~{\bar l}^c_3$&0&0&${1\over2}$&0&${1\over2}$&${1\over2}$&0&${3\over2}$&1\\
$\bar f_5,~{\bar l}^c_5$&0&$-{1\over2}$&0&0&0&${1\over2}$&0
&${3\over2}$&$-{1\over2}$\\
$h_1$&1&0&0&0&0&$-3$&1&1&0\\
$\bar h_1$&$-1$&0&0&0&0&3&$-1$&$-1$&0\\
$h_2$&0&1&0&0&0&$-1$&0&$-3$&1\\
$\bar h_2$&0&$-1$&0&0&0&1&0&3&$-1$\\
$h_3$&0&0&1&0&0&2&1&1&1\\
$\bar h_3$&0&0&$-1$&0&0&$-2$&$-1$&$-1$&$-1$\\
$h_{45}$&$-{1\over2}$&$-{1\over2}$&0&0&0&2&$-{1\over2}$&1&$-{1\over2}$\\
$\bar h_{45}$&${1\over2}$&${1\over2}$&0&0&0&$-2$&${1\over2}$&$-1$&${1\over2}$\\
\hline
\end{tabular}
\end{center}
\end{table}

\begin{table}[htp]
\caption{The singlet fields and their transformation properties
under $G_{\rm U(1)}$, $\rm U_A$, and three orthogonal linear combinations of
interest ($\rm U',U'',U'''$).}
\label{Table2}
\smallskip
\begin{center}
\footnotesize
\begin{tabular}{|l|rrrrr|rrrr|}\hline
&$\rm U_1$&$\rm U_2$&$\rm U_3$&$\rm U_4$&$\rm U_5$&$\rm U_A$
&$\rm U'$&$\rm U''$&$\rm U'''$\\
\hline
$\Phi_{12}$&$-1$&1&0&0&0&2&$-1$&$-4$&1\\
$\bar\Phi_{12}$&1&$-1$&0&0&0&$-2$&1&4&$-1$\\
$\Phi_{23}$&0&$-1$&1&0&0&3&1&4&0\\
$\bar\Phi_{23}$&0&1&$-1$&0&0&$-3$&$-1$&$-4$&0\\
$\Phi_{31}$&1&0&$-1$&0&0&$-5$&0&0&$-1$\\
$\bar\Phi_{31}$&$-1$&0&1&0&0&5&0&0&1\\
$\phi_{45}$&${1\over2}$&${1\over2}$&1&0&0&0&${3\over2}$&2&${3\over2}$\\
$\bar\phi_{45}$&$-{1\over2}$&$-{1\over2}$&$-1$&0&0&0&$-{3\over2}$
&$-2$&$-{3\over2}$\\
$\phi^+$&${1\over2}$&$-{1\over2}$&0&0&1&$-2$&$-{1\over2}$&4&${1\over2}$\\
$\bar\phi^+$&$-{1\over2}$&${1\over2}$&0&0&$-1$&2&${1\over2}$&$-4$
&$-{1\over2}$\\$\phi^-$&${1\over2}$&$-{1\over2}$&0&0&$-1$&0&${3\over2}$&0
&$-{3\over2}$\\
$\bar\phi^-$&$-{1\over2}$&${1\over2}$&0&0&1&0&$-{3\over2}$&0&${3\over2}$\\
$\phi_{3,4}$&${1\over2}$&$-{1\over2}$&0&0&0&$-1$&${1\over2}$&2&$-{1\over2}$\\
$\bar\phi_{3,4}$&$-{1\over2}$&${1\over2}$&0&0&0&1&$-{1\over2}$&$-2$&$1\over2$\\
$\eta_{1,2}$&0&0&0&1&0&0&0&0&0\\
$\bar\eta_{1,2}$&0&0&0&$-1$&0&0&0&0&0\\
$\Phi_{0,1,3,5}$&0&0&0&0&0&0&0&0&0\\
\hline
\end{tabular}
\end{center}
\end{table}

\begin{table}[htp]
\caption{The hidden SO(10) decaplets ({\bf10}) $T_i$ fields and their
transformation properties under $G_{\rm U(1)}$. In addition, we present their charges
under $\rm U_A$ and three orthogonal linear combinations of interest ($\rm
U',U'',U'''$).}
\label{Table3}
\footnotesize
\begin{center}
\begin{tabular}{|l|rrrrr|rrrr|}\hline
&$\rm U_1$&$\rm U_2$&$\rm U_3$&$\rm U_4$&$\rm U_5$&$\rm U_A$
&$\rm U'$&$\rm U''$&$\rm U'''$\\
\hline
$T_1$&$-{1\over2}$&0&${1\over2}$&0&0&${5\over2}$&0&0&${1\over2}$\\
$T_2$&$-{1\over2}$&$-{1\over2}$&0&0&$-{1\over2}$&${5\over2}$&0&0&$-1$\\
$T_3$&$-{1\over2}$&0&${1\over2}$&0&0&${5\over2}$&0&0&${1\over2}$\\
\hline
\end{tabular}
\vspace{2cm}
\caption{The hidden SU(4) fields and their transformation properties
  under $G_{\rm U(1)}$. $D_i$ represent sixplets ({\bf6}), whereas
($\widetilde F_i, \widetilde{\bar F}_i$) represent tetraplets (${\bf4},\bar{\bf4}$).
Moreover, we present the charges under $\rm U_A$ and three orthogonal linear
combinations of interest ($\rm U',U'',U'''$).}
\label{Table4}
\footnotesize
\medskip
\begin{tabular}{|l|rrrrr|rrrr|}\hline
&$\rm U_1$&$\rm U_2$&$\rm U_3$&$\rm U_4$&$\rm U_5$&$\rm U_A$
&$\rm U'$&$\rm U''$&$\rm U'''$\\
\hline
$D_1$&0&$-{1\over2}$&${1\over2}$&${1\over2}$&0&${3\over2}$&${1\over2}$&2&0\\
$D_2$&0&$-{1\over2}$&${1\over2}$&$-{1\over2}$&0&${3\over2}$&${1\over2}$&2&0\\
$D_3$&$-{1\over2}$&0&${1\over2}$&0&0&${5\over2}$&0&0&${1\over2}$\\
$D_4$&$-{1\over2}$&$-{1\over2}$&0&0&${3\over2}$&${3\over2}$&$-1$&2&0\\
$D_5$&0&$-{1\over2}$&${1\over2}$&0&0&${3\over2}$&${1\over2}$&2&0\\
$D_6$&0&${1\over2}$&$-{1\over2}$&0&0&$-{3\over2}$&$-{1\over2}$&$-2$&0\\
$D_7$&${1\over2}$&0&$-{1\over2}$&0&0&$-{5\over2}$&0&0&$-{1\over2}$\\
$\widetilde F_1$&$-{1\over4}$&${1\over4}$&$-{1\over4}$&0&$-{1\over2}$
&${1\over2}$&0&$-{9\over4}$&$-{1\over2}$\\
$\widetilde F_2$&${1\over4}$&${1\over4}$&$-{1\over4}$&0&${1\over2}$
&$-2$&$-{1\over2}$&${1\over4}$&${1\over2}$\\
$\widetilde F_3$&${1\over4}$&$-{1\over4}$&$-{1\over4}$&0&${1\over2}$
&$-{3\over2}$&$-{1\over2}$&${7\over4}$&0\\
$\widetilde F_4$&$-{1\over4}$&${3\over4}$&${1\over4}$&0&0
&${1\over2}$&0&$-{9\over4}$&1\\
$\widetilde F_5$&$-{1\over4}$&${1\over4}$&$-{1\over4}$&0&${1\over2}$
&$-{1\over2}$&$-1$&$-{1\over4}$&${1\over2}$\\
$\widetilde F_6$&$-{1\over4}$&${1\over4}$&$-{1\over4}$&0&$-{1\over2}$
&${1\over2}$&0&$-{9\over4}$&$-{1\over2}$\\
$\widetilde{\bar F}_1$
&$-{1\over4}$&${1\over4}$&${1\over4}$&${1\over2}$&$-{1\over2}$
&${3\over2}$&${1\over2}$&$-{7\over4}$&0\\
$\widetilde{\bar F}_2$
&$-{1\over4}$&${1\over4}$&${1\over4}$&$-{1\over2}$&$-{1\over2}$
&${3\over2}$&${1\over2}$&$-{7\over4}$&0\\
$\widetilde{\bar F}_3$
&${1\over4}$&$-{1\over4}$&${1\over4}$&0&$-{1\over2}$
&${1\over2}$&1&${1\over4}$&$-{1\over2}$\\
$\widetilde{\bar F}_4$
&$-{1\over4}$&${1\over4}$&${1\over4}$&0&$-{1\over2}$
&${3\over2}$&${1\over2}$&$-{7\over4}$&0\\
$\widetilde{\bar F}_5$
&$-{1\over4}$&$-{1\over4}$&${1\over4}$&0&$-{1\over2}$
&2&${1\over2}$&$-{1\over4}$&$-{1\over2}$\\
$\widetilde{\bar F}_6$
&$-{3\over4}$&${1\over4}$&$-{1\over4}$&0&0
&${3\over2}$&$-1$&$-{7\over4}$&0\\
\hline
\end{tabular}
\end{center}
\end{table}

\subsection{The First Scenario}
\label{sec:first}
The $\rm U_4(1)$ gauge symmetry is traceless (anomaly-free), and hence does not participate in the $\rm U_A(1)$ cancellation mechanism (unbroken). Specifically, $\rm U_4(1)$ is leptophobic since the leptons $\bar f_{2,3,5}$ and ${\bar l}^c_{2,3,5}$ are not charged under it from Table~\ref{Table1}. It is however interesting to note that $F_0$ and $F_1$ are indeed charged under $\rm U_4(1)$. The $\rm U_4(1)$ and $\rm U_Y(1)$ do not mix though: the Higgs doublets, which break the electroweak symmetry, are neutral under $\rm U_4$ (see $h_i,\bar h_i$ in
Table~\ref{Table1}). The mixing via gauge kinetic functions cannot be realized due to $\rm Tr\,(YU_4)=0$. This factor ``protects" the leptophobia, as otherwise the leptons would experience their $\rm U_4$ charges shifted away from zero.

Under the assumption that $F_0$ and $F_1$ contain the first two generations of the SM quarks, the $\rm U_4$ can remain unbroken during the $SU(5)\times U(1)_X$ symmetry breaking  at the usual GUT scale. The $\rm U_4$ symmetry may be broken radiatively at low energy if the singlet
fields $\eta_{1,2}$ and $\bar\eta_{1,2}$, which solely carry the $\rm U_4$ charges (see Table~\ref{Table2}), acquire suitable dynamical VEVs.
Although this $Z'$ can explain the dijet excess, it cannot explain the $WW$ excess since it clearly does not mix with the $Z$ boson. However, the top quark Yukawa coupling
is forbidden by the $U(1)_{\rm LP}$ gauge symmetry.

\subsection{The Second Scenario}
\label{sec:second}
Three linear combinations of $\rm U_{1,2,3,5}$ are orthogonal to $\rm U_A=U_1-3U_2+U_3+2U_5$ and traceless. Without loss of generality, we can choose the following basis: $\rm U'_1=U_3+2U_5$, $\rm U'_2=U_1-3U_2$, $\rm U'_3=3U_1+U_2+4U_3-2U_5$. Since the leptons transform as $\bar f_{2,5},\ell^c_{2,5}:(0,{3\over2},-{1\over2})$; $\bar f_3,\ell^c_3:({3\over2},0,1)$ under $U'_i$, there is a unique leptophobic linear combination of $U'_i$: $\rm U'\propto 2U'_1-U'_2-3U'_3 \propto U_1+U_3-U_5$. The $U'$ gauge symmetry is by construction anomaly-free and leptophobic, and some of the Higgs pentaplets are charged under it ({\it i.e.}, mixed). The charges of all fields under $\rm U'$ are given in the Tables, along with two extra traceless combinations which can be chosen as $\rm U''=U_1-3U_2+U_3+2U_5$ and $\rm U'''=U_2+U_3+U_5$. From the tables we find that only a very limited set of fields is neutral under $\rm U'$
\begin{equation}
F_2,\bar F_5,\Phi_{31},\bar\Phi_{31},T_1,T_2,T_3,D_3,D_7\ ,
\label{eq:vevs}
\end{equation}
and therefore their VEVs will not break the $\rm U'$ gauge symmetry. The challenge is whether the usual D- and F-flatness conditions can be satisfied with such a limited set of VEVs since it generally breaks the hidden sector gauge groups. This problem may be solved if one introduces
the non-renormalizable superpotential~\cite{Lopez:1996ta}.

It can be verified that if $\rm Tr\,(YU')=0$, then $\rm U'$ can indeed remain unbroken down to low energy, thus permitting the $\rm U'$ charges to remain unshifted and the leptophobia protected. Moreover, only $F_2$ and $\bar F_5$ can break the $SU(5)\times U(1)_X$ gauge symmetry since they are not charged under $\rm U'$. Unlike
the previous studies~\cite{search} where $F_4$ contains the third-generation quarks,
we consider $F_0$, $F_1$, and $F_3$ as the first, second and third generations, respectively. 
Additionally, $F_4$ and $\bar F_4$ can form vector-like particles at the intermediate scale such that string-scale gauge coupling unification can be achieved~\cite{goodies}.
For the pentaplet Higgs fields $h_i$ and ${\bar h}_i$, for simplicity, we assume that $h_2$ and ${\bar h}_2$
are vector-like and have mass around the usual GUT or string scale. Moreover,
the triplets in the ($h_1$, ${\bar h}_1$) and ($h_3$, ${\bar h}_3$)
will be light at low energy since they are charged under $U(1)_{\rm LP}$. Therefore, in the low energy supersymmetric SM, there will be two pairs of Higgs doublets and two pairs of Higgs triplets. In particular,
$H_d$ is a linear combination of $h_1$ and $h_3$,
while $H_u$ is a linear combination of ${\bar h}_1$ and
${\bar h}_3$ and its dominant component is ${\bar h}_3$. Notably,
($h_1$, ${\bar h}_1$) and ($h_3$, ${\bar h}_3$) are
charged under $U(1)_{\rm LP}$, and then $Z$ and $Z'$ are mixed after the Higgs fields acquire the VEVs.
Given this scenario, we could explain the diboson and dijet excesses~\cite{Hisano:2015gna}.
In particular, unlike the first scenario,
the top quark Yukawa coupling $F_3 {\bar f}_3 {\bar h}_3$
is allowed by the $U(1)_{\rm LP}$ gauge symmetry. And the down-type quark Yukawa couplings
such as $F_0 F_0 h_1$, $F_0 F_1 h_1$ and $F_1 F_1 h_1$ can be realized
at renormalizable level. While all the other SM fermion Yukawa couplings should be generated via
high-dimensional operators. Furthermore, for simplicity, we assume that all the Higgs fields
except the SM-like Higgs field are heavy and then are still undetected at the LHC.

\section{Leptophobic Z$'$ from $SU(4)_C \times SU(2)_L \times SU(2)_R \times U(1)_X$ on D-branes}

A phenomenologically interesting intersecting D-brane model has been studied in Refs.~\cite{Chen:2007px,Chen:2007zu}. A variation of this model with a different hidden sector was also studied in Refs.~\cite{Maxin:2011ne,Chen:2007ms}. The full gauge symmetry of the model is given by $[{\rm U}(4)_C \times {\rm U}(2)_L \times {\rm U}(2)_R]_{\rm observable} \times [ {\rm U}(2) \times {\rm USp}(2)^2]_{\rm hidden}$, with the matter content shown in Tables~\ref{Spectrum} and~\ref{VectorlikeSpectrum}.  Note that in Table~\ref{Spectrum}, $a$, $b$, $c$, etc. refer to different stacks of D-branes which wrap cycles of the compactified manifold and which generically intersect at angles.  A stack of $2N$ D-branes results in a $U(N)$ gauge group in the world-volume of each stack.  Strings localized at the intersection between two stacks result in massless fermions in the bifundamental representation of the gauge group of each stack. Vector-like matter may also be present between stacks which do not intersect, again in the bifundamental representation of each stack's gauge group.   

\begin{table}
[t] \footnotesize
\renewcommand{\arraystretch}{1.0}
\caption{The chiral superfields, their multiplicities
and quantum numbers under the gauge symmetry $[{\rm U}(4)_C \times {\rm U}(2)_L \times {\rm
U}(2)_R]_{\rm observable} \times [ {\rm U}(2) \times {\rm USp}(2)^2]_{\rm hidden}$, where
$Q_X = Q_a+ 2(Q_{b}+Q_{c}+3Q_d)$. Here $a$, $b$, $c$, etc. refer to different stacks of D-branes.}
\label{Spectrum}
\begin{center}
\begin{tabular}{|c||c|c||c|c|c|c|c|c|c|c|c||c|}\hline
& Mult. & Quantum Number & $Q_a$ & $Q_b$ & $Q_c$ & $Q_d$ & $Q_X$ & Field
\\
\hline\hline
$ab$ & 3 & $(4,\overline{2},1,1,1,1)$ & \ 1 & $-1$ & \ 0 & 0 & $-1$ &
$F_L(Q_L, L_L)$\\
$ac$ & 3 & $(\overline{4},1,2,1,1,1)$ & $-1$ & 0 & \ 1  & 0 & \ 1 &
$F_R(Q_R, L_R)$\\
\hline
$bd$ & 1 & $(1,\overline{2},1,2,1,1)$ & 0 & $-1$ & 0  & 1 & $4$ & $X_{bd}$\\
$cd$ & 1 & $(1,1,2,\overline{2},1,1)$ & 0 & $0$ & 1 & -1 & $-4$ & $X_{cd}$\\
\hline
$b4$ & 3 & $(1,\overline{2},1,1,1,2)$ & 0 & $-1$ & 0 & 0 & $-2$ &
$X_{b3}^i$ \\
$c3$ & 3 & $(1,1,2,1,\overline{2},1)$ & 0 & 0 & \ 1  & 0 & \ 2 &  
$X_{c3}^i$
\\
$d3$ & 1 & $(1,1,1,\overline{2},2,1)$ & 0 & $0$ & 1  & -1 & $-6$  & $X_{cd}$\\
$d4$ & 1 & $(1,1,1,2,1,\overline{2})$ & 0 & $0$ & 1  &  1 & $6$  & $X_{cd}$\\
$b_{S}$ & 2 & $(1,3,1,1,1,1)$ & 0 & \ 2 & 0    & 0 & \ 4 &  $T_L^i$ \\
$b_{A}$ & 2 & $(1,\overline{1},1,1,1,1)$ & 0 & $-2$ & 0 & 0 & $-4$ &
$S_L^i$
\\
$c_{S}$ & 2 & $(1,1,\overline{3},1,1,1)$ & 0 & 0 & 0 & $-2$  & $-4$ &
$T_R^i$
\\
$c_{A}$ & 2 & $(1,1,1,1,1,1)$ & 0 & 0 & \ 2  & 0 & 4 & $S_R^i$ \\
\hline
\end{tabular}
\end{center}
\end{table}

\begin{table}
[t] \footnotesize
\renewcommand{\arraystretch}{1.0}
\caption{The vectorlike superfields, their multiplicities
and quantum numbers under the gauge symmetry $[{\rm U}(4)_C \times {\rm U}(2)_L \times {\rm
U}(2)_R]_{\rm observable} \times [ {\rm U}(2) \times {\rm USp}(2)^2]_{\rm hidden}$, where
$Q_X = Q_a+ 2(Q_{b}+Q_{c}+3Q_d)$. Here $a$, $b$, $c$, etc. refer to different stacks of D-branes.}
\label{VectorlikeSpectrum}
\begin{center}
\begin{tabular}{|c||c|c||c|c|c|c|c|c|c|c|c||c|}\hline
& Mult. & Quantum Number & $Q_a$ & $Q_b$ & $Q_c$ & $Q_d$ & $Q_X$ & Field
\\
\hline\hline
$ab'$ & 3 & $(4,2,1,1,1,1)$ & \ 1 & \ 1 & 0  & 0 & \ 3 & $\Omega^i_L$ \\
& 3 & $(\overline{4},\overline{2},1,1,1,1)$ & $-1$ & $-1$ & 0 & 0 & $-3$ &
$\overline{\Omega}^i_L$ \\
\hline
$ac'$ & 3 & $(4,1,2,1,1,1)$ & \ 1 & 0 & \ 1  & 0 &  \ 3 & $\Phi_i$ \\
& 3 & $(\overline{4}, 1, \overline{2},1,1,1)$ & $-1$ & 0 & $-1$ & 0 & $-3$
& $\overline{\Phi}_i$\\
\hline
$ad$ & 2 & $(4,1,1,\overline{2},1,1)$ & \ 1 & 0 & 0 & -1 & \ -5 &  $\varphi_i$\\
& 2 & $(\overline{4},1,1,2,1,1)$ & $-1$ & 0 & 0 &  $1$ & $5$    &  $\overline{\varphi}_i$\\ 
\hline
$ad'$ & 1 & $(4,1,1,2,1,1)$ & \ 1 & 0 & 0 & 1 & \ 7 & $\varsigma$ \\
& 1 & $(\overline{4},1,1,\overline{2},1,1)$ & $-1$ & 0 & 0 &  $-1$ & $-7$& $\overline{\varsigma}$\\
\hline
$bc$ & 6 & $(1,2,\overline{2},1,1,1)$ & 0 & 1 & $-1$  & \ 0 & 0 &
$H^i\left(H_u, H_d\right)$\\
     & 6  & $(1,\overline{2},2,1,1,1)$ & 0 & $-1$ & 1 & 0  & \ 0 & $\overline{H}^i\left(\overline{H}_u, \overline{H}_d\right)$\\
\hline		
$bc'$ & 1 & $(1,2,2,1,1,1)$ & 0 & 1 & 1 & 0 & \ 4& $\mathcal{H}\left(\mathcal{H}_u, \mathcal{H}_d\right)$
\\ 
     & 1  & $(1,\overline{2},\overline{2},1,1,1)$ & 0 & $-1$ & $-1$  & 0 & -4 &	$\overline{\mathcal{H}}\left(\overline{\mathcal{H}}_u, \overline{\mathcal{H}}_d\right)$
\\
\hline
$bd'$ & 1 & $(1,2,1,2,1,1)$ & 0 & 1 & 0 & 1 & \ 8 & $\xi$ \\
     & 1  & $(1,\overline{2},1,\overline{2},1,1)$ & 0 & 0 & -1  & -1 & -8 &	$\overline{\xi}$\\
\hline
	$cd'$ & 1 & $(1,1,2,2,1,1)$ & 0 & 1 & 0 & 1 & \ 8 & $\psi$ \\
     & 1  & $(1,1,\overline{2},\overline{2},1,1)$ & 0 & 0 & -1  & -1 & -8 &	$\overline{\psi}$\\	
\hline		
\hline
\end{tabular}
\end{center}
\end{table}

Since ${\rm U}(N) = {\rm SU}(N)\times {\rm U}(1)$, associated with each of the stacks $a$, $b$, $c$, and $d$ are ${\rm U(1)}$ gauge groups,
denoted as ${\rm U}(1)_a$, ${\rm U}(1)_b$, ${\rm U(1)}_c$, and ${\rm U(1)}_d$. In general, these U(1)s are anomalous.  The anomalies associated with these U(1)s are canceled by a generalized Green-Schwarz (G-S) mechanism that involves untwisted R-R forms. As a result, the gauge bosons of these Abelian groups generically become massive. The G-S couplings determine the exact linear combinations of U(1) gauge bosons that become massive.  Some linear combinations may remain massless if certain conditions are satisfied.  

\begin{table}
[htb] \footnotesize
\renewcommand{\arraystretch}{1.0}
\caption{The chiral superfields, their multiplicities
and quantum numbers under the gauge symmetry $[{\rm SU}(3)_C \times {\rm SU}(2)_L \times {\rm
U}(1)_Y \times {\rm U}(1)_B ]_{\rm observable} \times [ {\rm SU}(2) \times {\rm USp}(2)^2]_{\rm hidden}$.}
\label{MSSMSpectrum}
\begin{center}
\begin{tabular}{|c||c|c||c|c|c|c|c|c|c||c|}\hline
& Mult. & Quantum Number  & $Q_{I3R}$ & $Q_{B-L}$ & $Q_{3B+L}$ & $Q_{Y}$ & $Q_B$ & Field
\\
\hline\hline
$a1b$ & 3 & $(3,\overline{2},1,1,1,1,1)$ & 0 & 1/3 & 1 & 1/6 & 1/3 & $Q_L$\\
$a1c2$ & 3 & $(\overline{3},1,1,1,1,1,1)$ & -1/2 & -1/3 & -1  & -2/3 & -1/3 & $U_R$\\
$a1c1$ & 3 & $(\overline{3},1,1,1,1,1,1)$ & 1/2 & -1/3 & -1  & 1/3 & -1/3 & $D_R$\\
$a2b$ & 3 & $(1,\overline{2},1,1,1,1,1)$ & 0 & -1 & 1 & -1/2 &  0 & $L$\\
$a2c1$ & 3 & $(1,2,1,1,1,1,1)$ & 1/2 & 1 & -1  & 1 & 0 & $E_R$\\
$a2c2$ & 3 & $(1,2,1,1,1,1,1)$ & -1/2 & 1 & -1  & 0 & 0 & $N_R$\\
\hline
\hline
$bc1$ & 6 & $(1,2,1,1,1,1,1)$ & -1/2 & 0 & 0  & -1/2 & 0 &  $H_d^i$\\
$bc2$ & 6 & $(1,2,1,1,1,1,1)$ & 1/2 & 0 & 0  & 1/2 & 0 &  $H_u^i$\\
$bc1'$ & 1 & $(1,\overline{2},1,1,1,1,1)$ & -1/2 & 0 & 4  & -1/2 & 1 &  $\mathcal{H}_d$\\ 
$bc2'$ & 1 & $(1,\overline{2},1,1,1,1,1)$ &  1/2 & 0 & 4  &  1/2 & 1 &  $\mathcal{H}_u$\\ 
\hline
\end{tabular}
\end{center}
\end{table}

\begin{table}
[htb] \footnotesize
\renewcommand{\arraystretch}{1.0}
\caption{The chiral hidden sector superfields, their multiplicities
and quantum numbers under the gauge symmetry $[{\rm SU}(3)_C \times {\rm SU}(2)_L \times {\rm
U}(1)_Y \times {\rm U}(1)_B ]_{\rm observable} \times [ {\rm SU}(2) \times {\rm USp}(2)^2]_{\rm hidden}$.}
\label{HiddenSpectrum}
\begin{center}
\begin{tabular}{|c||c|c||c|c|c|c|c|c|c||c|}\hline
& Mult. & Quantum Number  & $Q_{I3R}$ & $Q_{B-L}$ & $Q_{3B+L}$ & $Q_{Y}$ & $Q_B$ & Field
\\
\hline\hline
$bd$ & 1 & $(1,\overline{2},1,2,1,1,1)$ & 0 & 0 & -4  & 0 & -1 & $X_{bd}$\\
$c1d$ & 1 & $(1,1,1,\overline{2},1,1,1)$ & 1/2 & 0 & 4 & 1/2 & $1$ & $X_{c1d}$\\
$c2d$ & 1 & $(1,1,1,\overline{2},1,1,1)$ & -1/2 & 0 & 4 & -1/2 & $1$ & $X_{c2d}$\\
$b4$ & 3 & $(1,\overline{2},1,1,1,2,1)$ & 0 & 0 & -2 & 0 & -1/2 &$X_{b3}^i$ \\
$c13$ & 3 & $(1,1,1,1,1,\overline{2},1)$ & 1/2 & 0 & -2  &  1/2 & -1/2 &  $X_{c13}^i$\\
$c23$ & 3 & $(1,1,1,1,1,\overline{2},1)$ & -1/2 & 0 & -2  &  -1/2 & -1/2 &  $X_{c23}^i$\\
$d3$ & 1 & $(1,1,1,1,\overline{2},2,1)$ & 0 & 0 & -6  & 0 & -3/2 & $X_{cd}$\\
$d4$ & 1 & $(1,1,1,1,2,1,\overline{2})$ & 0 & 0 & -6  & 0 & -3/2  & $X_{cd}$\\
$b_{S}$ & 2 & $(1,3,1,1,1,1,1)$ & 0 & 0 & -4  & 0 & -1 &  $T_L^i$ \\
$b_{A}$ & 2 & $(1,1,1,1,1,1,1,1)$ & 0 & 0 & 4  & 0 & 1 & $S_L^i$\\
$c_{S}$ & 2 & $(1,1,1,1,1,1)$ & 0 & 0 & -4 & 0  & -1 &$T_R^i$\\
\hline
\end{tabular}
\end{center}
\end{table}

As shown in Ref.~\cite{Maxin:2011ne}, precisely one linear combination of the present
model remains massless and anomaly-free:
\begin{equation}
{\rm U}(1)_X = {\rm U}(1)_a + 2\left[{\rm U}(1)_b + {\rm U}(1)_c + 3{\rm U}(1)_d
\right].
\end{equation}
Thus, the effective gauge symmetry of the model at the string scale is given by
\begin{equation}
{\rm SU}(4)_C \times {\rm SU}(2)_L \times {\rm SU}(2)_R \times {\rm
U}(1)_X \times \left[{\rm SU}(2) \times {\rm USp}(2)^2\right].
\end{equation}
As can be seen from Table~\ref{Spectrum}, the superfields $F_L^i(Q_L,L_L)$ carry charge $Q_X = -1$, the superfields $F_R^i(Q_R, L_R)$ carry
charge $Q_X = +1$. In addition, there are the Higgs superfields $H_u^i$, $H_d^i$ in the $bc$ sector which are uncharged under U(1)$_X$ while the Higgs superfields $\mathcal{H}_u$, $\mathcal{H}_d$ in the $bc'$ sector carry charges $Q_X = \pm 4$ respectively.  

It should be noted that the Yukawa couplings with the Higgs superfields $H_u^i$, $H_d^i$ are allowed by the global $U(1)$ charges.
The resulting Yukawa mass matrices
for quarks and leptons are of rank 3, and it has been shown that it is possible to obtain the correct masses and mixings for all quarks and leptons~\cite{Chen:2007zu,Chen:2007px}.  On the other hand, the Yukawa couplings with the Higgs superfields from the $bc'$ sector $\mathcal{H}_u$, $\mathcal{H}_d$ are forbidden.  
In addition we may form a $\mu$-term in the superpotential of the form
In addition we may form a $\mu$-term in the superpotential of the form
\begin{equation}
\label{muterm}
W_{\mu} = \frac{y^{ijkl}}{M_{St}}S_L^i S_R^j H^k_u H^l_d,  
\end{equation}
which is TeV-scale, Where $S_R^j$ receive string scale VEVs, $M_{St}$ is the string scale, and the VEVs of $S_L^i$ are 
TeV-scale.  The $\mu$-term may be fine-tuned so that only a pair 
of Higgs eigenstates $H_u$ and $H_d$ remain light, as in the MSSM.

The gauge symmetry is first broken by splitting the D-branes as $a\rightarrow a1 + a2$ with $N_{a1}=6$ and $N_{a2}=2$, and $c\rightarrow c1 + c2$ with $N_{c1}=2$ and $N_{c2}=2$. After splitting the D6-branes, the gauge symmetry of the observable sector is 
\begin{equation}
SU(3)_C \times SU(2)_L \times U(1)_{I3R} \times U(1)_{B-L} \times U(1)_{3B+L},
\end{equation}
where 
\begin{eqnarray}
U(1)_{I3R} = \frac{1}{2}(U(1)_{c1} - U(1)_{c2}), \ \ \  
U(1)_{B-L} = \frac{1}{3}(U(1)_{a1} - 3U(1)_{a2}), 
\end{eqnarray}
and
\begin{equation}
U(1)_{3B+L}= -[U(1)_{a1}+U(1)_{a2} + 2(U(1)_{b}+U(1)_{c1}+U(1)_{c2}+3U(1)_{d})],
\end{equation}
and ${\rm U}(1)_{3B+L} = -{\rm U}(1)_X$.

The gauge symmetry must be further broken to the SM, with the possibility of one or more additional U(1) gauge symmetries. In particular, 
the $U(1)_{B-L}\times U(1)_{I_{3R}}\times U(1)_{3B+L}$ gauge symmetry may be broken by assigning VEVs to the right-handed neutrino fields $N_R^i$.  In this case, the gauge symmetry is broken to
\begin{equation}
\left[SU(3)_C \times SU(2)_L \times U(1)_Y \times U(1)_B\right]_{observable} \times \left[SU(2)\times USp(2)^2\right]_{hidden}
\label{MSSM_B}
\end{equation}
where 
\begin{eqnarray}
&U(1)_Y = \frac{1}{6}\left[U(1)_{a1}-3U(1)_{a2}+3U(1)_{c1}-3U(1)_{c2}\right] \\ \nonumber
& = \frac{1}{2}U(1)_{B-L}+U(1)_{I3R}.
\end{eqnarray}
and
\begin{eqnarray}
&U(1)_B = \frac{1}{4}[U(1)_{B-L} + U(1)_{3B+L}]\\ \nonumber
&= -[\frac{1}{6}U(1)_{a1}+\frac{1}{2}(U(1)_{a2} + U(1)_{b}+U(1)_{c1}+U(1)_{c2}+3U(1)_{d})].
\end{eqnarray}

We will assume that all exotic matter, shown in Table~\ref{HiddenSpectrum}, may become massive, as shown in Ref.~\cite{Chen:2007ms}. The resulting low-energy field content is shown in Tables~\ref{MSSMSpectrum} and along with their charges under $U(1)_{I3R}$, $U(1)_{B-L}$, $U(1)_{3B+L}$, $U(1)_Y$, and $U(1)_B$. Note that the quarks are charged under $U(1)_B$, but the leptons are not.      

The extra gauge symmetry $U(1)_B$ may then be spontaneously broken if the SM singlet fields $S^i_L$, which carry a charge of $+1$ under $U(1)_B$ obtain VEVs at some scale $\Lambda =\mathcal{O}({\rm TeV})$. Thus, the model may possess a leptophobic $Z'$ boson with an $\mathcal{O}({\rm TeV})$ mass.  In order to explain the diboson excess observed by ATLAS, we will take this scale to be $2-3$~TeV.   

The electroweak symmetry is broken when some of the Higgs fields obtain VEVs.  In order to obtain masses and mixings for the quarks and leptons, we will assume that the dominate VEVs are acquired by the Higgs fields $H_u^i, H_d^i$ in the $bc$ sector which are uncharged under $U(1)_B$. As noted previously, the Yukawa couplings for quarks and leptons via these Higgs fields are present, and realistic masses and mixings may be obtained. The Yukawa couplings with the Higgs fields $\mathcal{H}_u, \mathcal{H}_d$ from the $bc'$ sector, which carry charges of $\pm 1$ under $U(1)_B\equiv U(1)_{\rm LP}$ respectively, are perturbatively forbidden by the global $U(1)$ charges.  We will assume that these fields may also obtain a subdominant VEV with respect to the Higgs fields in the $bc$ sector. The $Z$ and $Z'$ bosons will then be mixed as a result. Thus, we might explain the diboson and dijet excesses~\cite{Hisano:2015gna}. 

Clearly, the requirement that a Higgs field be charged under the leptophobic $U(1)$ in order to obtain mixing between the $Z$ and $Z'$ bosons 
results in the Yukawa couplings with this Higgs field being forbidden, which seems to be a generic problem for models of this type. 
In the present context, this leads to the requirement of an extended Higgs sector with some Higgs fields charged under $U(1)_B$ and some which 
are not for which the Yukawa couplings are present. Specifically, the Yukawa couplings with Higgs fields in the $bc$ sector are allowed by the global
$U(1)$ charges carried by these fields:
\begin{equation}
W_Y = y^{ijk}_U H_u^i Q_L^j U_R^k + y^{ijk}_D H_d^i Q_L^j D_R^k + y^{ijk}_ {\nu} H_u^i L_L^j N_R^k + y^{ijk}_e H_d^i L^j E_R^k,
\end{equation}
while the Yukawa couplings with the extra Higgs fields $\mathcal{H}_u$ and $\mathcal{H}_d$ from the $bc'$ sector
which are charged under
$U(1)_B$ are perturbatively forbidden. 
We assume that these extra Higgs fields have masses so that they have not been observed at the LHC.  
We shall defer a detailed study of such extra Higgs bosons to later work.  

\section{Diboson Signals at the 8 TeV LHC}

The production of a leptophobic $Z'$ boson at the 8 TeV LHC is strongly constrained by three independent search regions: $Z' \rightarrow jj$~\cite{Brehmer:2015cia}, $Z' \rightarrow t \bar{t}$~\cite{Khachatryan:2015sma}, and $Z' \rightarrow Zh$~\cite{Khachatryan:2015bma}. Given the anomaly free nature of the $Z'$ boson described in the models presented here, the $Z' \rightarrow ZZ$ and $Z' \rightarrow Z \gamma$ are forbidden. Therefore, we only apply the following three constraints:
\begin{equation}
\sigma(pp \rightarrow Z') \times Br (Z' \rightarrow jj) \sim 91_{-45}^{+53}~{\rm fb} ~~~(1 \sigma~{\rm fit})
\label{lhc_constraint1}
\end{equation}
\begin{equation}
\sigma(pp \rightarrow Z') \times Br (Z' \rightarrow t \bar{t}) \lesssim 11 (18)~{\rm fb} ~~~(95\% ~{\rm CL})
\label{lhc_constraint2}
\end{equation}
\begin{equation}
\sigma(pp \rightarrow Z') \times Br (Z' \rightarrow Zh) \lesssim 7~{\rm fb} ~~~(95\% ~{\rm CL}) \ 
\label{lhc_constraint3}
\end{equation}
\noindent The $\sigma(pp \rightarrow Z') \times Br (Z' \rightarrow jj) \sim 91_{-45}^{+53}~{\rm fb}$ constraint consists of a $1 \sigma$ fitted cross-section~\cite{Brehmer:2015cia}, thus we also consider a more relaxed alternative. Softening these $1 \sigma$ boundaries, we shall also observe the result of applying only a 90\% CL upper bound of $\sigma(pp \rightarrow Z') \times Br (Z' \rightarrow jj) \lesssim 170~{\rm fb}$~\cite{Brehmer:2015cia}. The upper limit on $t \bar{t}$ resonances of 11 fb established by the CMS Experiment~\cite{Khachatryan:2015sma} corresponds to a decay width of 20 GeV for a 2 TeV $Z'$ boson, whereas the 18 fb upper limit correlates to a 200 GeV decay width, though we shall generally only regard the less stringent 18 fb constraint in this analysis when phenomenologically constraining the gauge coupling $g_{Z'}$.

\begin{figure}[htp]
        \centering
        \includegraphics[width=1.00\textwidth]{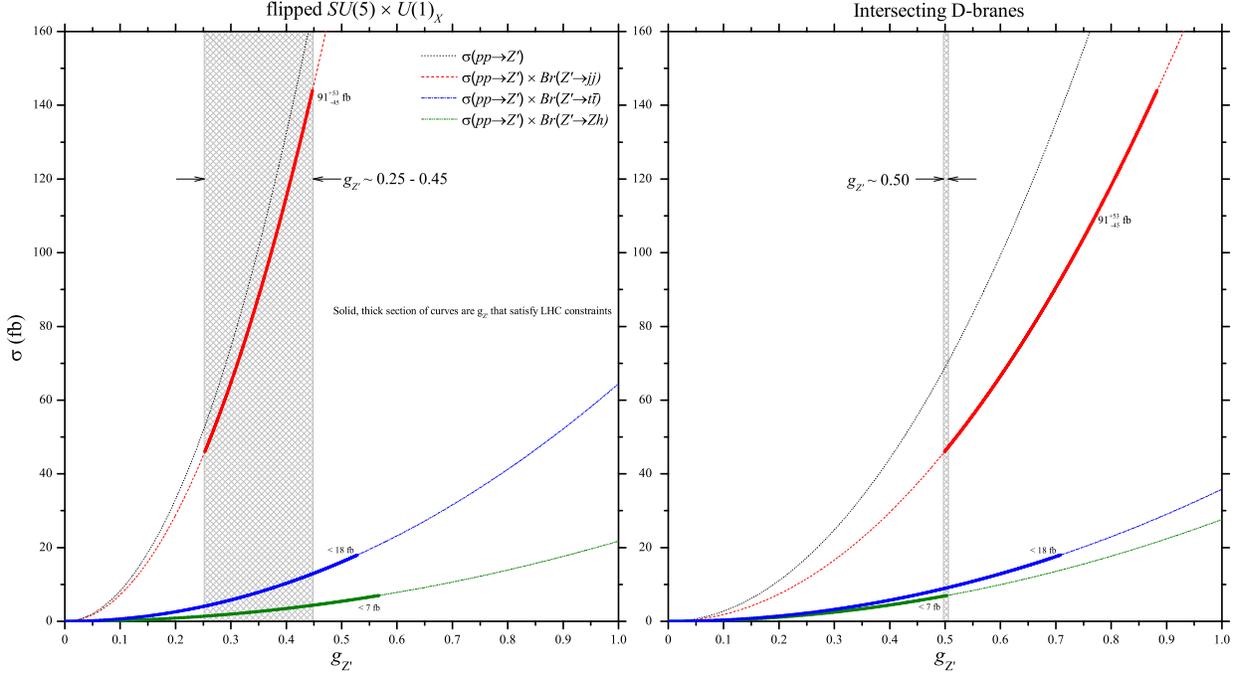}
        \caption{Depiction of the cross-section $\sigma$ as a function of the $Z'$ boson gauge coupling $g_{Z'}$ on $U(1)_{LP}$ for the flipped $SU(5) \times U(1)_X$ models (left frame) and the intersecting D-brane model (right frame). Here we only consider the LHC constraints $\sigma(pp \rightarrow Z') \times Br (Z' \rightarrow jj) \sim 91_{-45}^{+53}~{\rm fb}$, $\sigma(pp \rightarrow Z') \times Br (Z' \rightarrow Zh) \lesssim 7~{\rm fb}$, and the 200 GeV decay width $t \bar{t}$ constraint of $\sigma(pp \rightarrow Z') \times Br (Z' \rightarrow t \bar{t}) \lesssim 18~{\rm fb}$. Those $g_{Z'}$ satisfying these three constraints are shown as solid, thick sections of the curves, with the intersection marked by the cross-hatched region, where $g_{Z'} \sim 0.25 - 0.45$ for the flipped models and $g_{Z'} \sim 0.5$ for D-branes. If the LHC constraint on $Z' \rightarrow jj$ is relaxed to only an upper limit of $\sigma(pp \rightarrow Z') \times Br (Z' \rightarrow jj) \lesssim 170~{\rm fb}$, then loose model constraints of $g_{Z'} \lesssim 0.49$ for flipped models and $g_{Z'} \lesssim 0.50$ for D-branes are obtained.}
        \label{fig:gzp_plot}
\end{figure}

The calculation of the partial decay widths requires the quark and Higgs field charges on the leptophobic $U(1)_{\rm LP}$. The quark decay width is given by~\cite{Hisano:2015gna}

\begin{equation}
\Gamma(Z' \rightarrow q \bar{q}) = \frac{g_{Z'}^2 N_C}{24 \pi} M_{Z'} \left[ Q_{LP_{q_L}}^2 + Q_{LP_{q_R}}^2 - \left( Q_{LP_{q_L}} - Q_{LP_{q_R}} \right)^2 \left( \frac{m_q}{M_{Z'}} \right)^2  \right] \sqrt{1 - 4 \left( \frac{m_q}{M_{Z'}} \right)^2} 
\label{gamma_qq}
\end{equation}

\noindent where $g_{Z'}$ is the $U(1)_{LP}$ gauge coupling, $N_C = 3$ represents the number of colors, and $Q_{LP_{q_L}}$, $Q_{LP_{q_R}}$ are the left- and right-handed charges of the quark content on $U(1)_{LP}$. Here we take $M_{Z'} = 2$ TeV and $m_t = 174.4$ GeV~\cite{Tevatron:2014cka,Leggett:2014hha}. The $\Gamma(Z' \rightarrow W^+W^-)$ decay width can be computed from~\cite{Hisano:2015gna}

\begin{equation}
\Gamma(Z' \rightarrow W^+W^-) = \frac{g_{Z'}^2}{48 \pi} M_{Z'} Q_{LP_{H_u}}^2 \sin^4 \beta
\label{gamma_ww}
\end{equation}

\noindent with $Q_{LP_{H_u}}$ as the Higgs field charge on $U(1)_{LP}$, and $\beta$ the angle between the up and down Higgs VEVs. The equivalence theorem suggests for a heavy $Z'$ boson in the decoupling limit that

\begin{equation}
\Gamma(Z' \rightarrow Zh) = \Gamma(Z' \rightarrow W^+W^-)
\label{gamma_zh}
\end{equation}

\noindent indicating that the branching ratio for these two decay modes are equivalent. As a result, application of the strong upper limit $\sigma(pp \rightarrow Z') \times Br (Z' \rightarrow Zh) \lesssim 7~{\rm fb}$ likewise tightly constrains our $Z' \rightarrow W^+ W^-$ production as well. Proper normalization is required for accurate $U(1)_{LP}$ gauge coupling evolution, presumably achieving unification with the $SU(3)_C$, $SU(2)_L$, and $U(1)_Y$ gauge couplings at the GUT scale. For the flipped $SU(5) \times U(1)_X$ models, the explicit normalization factor is $\frac{1}{3}$. All massless fields have conformal dimension 1, generating the transformation $Q' \rightarrow Q' / \sqrt{3}$, where $Q'$ are the charges on $U'$ given in Tables~\ref{Table1}, \ref{Table2}, \ref{Table3}, and \ref{Table4}. Therefore, the $U'$ gauge coupling evolves according to the beta function $b' = \frac{1}{3} {\rm Tr} (Q')^2$. For convenience, this factor of $\frac{1}{3}$ in the decay widths is introduced directly into the calculation of the $\sigma(pp \rightarrow Z')$ cross-section in our numerical results via a normalization factor $n$ in Eq. (\ref{sigma_zp}), explicitly implementing $n = \frac{1}{3}$ for flipped models. For the intersecting D-brane model, an explicit normalization factor on $U(1)_B$ it not yet known, though it is expected that it is of unity order, hence, we assume any normalization of ${\cal O}(1)$ is already assimilated into the phenomenologically constrained value of the $U(1)_{B}$ gauge coupling $g_{Z'}$. We thus apply $n = 1$ in Eq. (\ref{sigma_zp}) for intersecting D-branes. An estimate of the 8 TeV LHC cross-section for $Z'$ boson production is given as~\cite{Leike:1998wr,Hisano:2015gna,Anchordoqui:2015uea}

\begin{equation}
\sigma(pp \rightarrow Z') \simeq n \times \left[ 5.2 \left( \frac{2 \Gamma ( Z' \rightarrow u \bar{u} ) + \Gamma ( Z' \rightarrow d \bar{d} )}{\rm GeV} \right) \right]{\rm fb}
\label{sigma_zp}
\end{equation}

\noindent using the decay widths given in Eq. (\ref{gamma_qq}).

The charges of the MSSM content on $U'$ are given in Table~\ref{Table1} for leptophobic flipped $SU(5) \times U(1)_X$ models. The matter fields $F_3$, $F_1$, and $F_4$ represent the first, second, and third generations, respectively, and ${\bar h_1}$ contains $H_u$. Thus, to compute the decay widths and branching ratios, the $U' = U(1)_{LP}$ charges used are $Q_{LP_{u_L}} = Q_{LP_{u_R}} = Q_{LP_{d_L}} = Q_{LP_{d_R}} = 1$, $Q_{LP_{c_L}} = Q_{LP_{c_R}} = Q_{LP_{s_L}} = Q_{LP_{s_R}} = -\frac{1}{2}$, $Q_{LP_{t_L}} = Q_{LP_{t_R}} = Q_{LP_{b_L}} = Q_{LP_{b_R}} = -\frac{1}{2}$, and $Q_{LP_{H_u}} = -1$. A review of Table~\ref{Table1} containing the MSSM content charges in the observable sector for the flipped $SU(5) \times U(1)_X$ model shows there are no electrically neutral particles at the electroweak scale charged on $U'$, hence there are no invisible decays. The only $Z'$ decay modes present then are $Z' \rightarrow jj$, $Z' \rightarrow t \bar{t}$, $Z' \rightarrow WW$, and $Z' \rightarrow Zh$. Though the No-Scale Supergravity boundary conditions at the unification scale are not applied here, we do implement a preferred No-Scale Supergravity angle $\beta$ using tan$\beta = 25$~\cite{Li:2013naa} . The decay widths were computed with both tan$\beta = 5$ and tan$\beta = 25$, resulting in only a mere $\sim 5 \%$ increase in the cross-section for the larger tan$\beta$, a safely negligible delta for our purposes here. Given this lack of significant variation in the cross-section as a function of $\beta$, there is essentially only one free-parameter remaining, the $Z'$ gauge coupling $g_{Z'}$. Therefore, we phenomenologically constrain the value of $g_{Z'}$ using the LHC constraints on $Z' \rightarrow jj$, $Z' \rightarrow t \bar{t}$, and $Z' \rightarrow Zh$. The branching ratios are independent of variation in $g_{Z'}$, with the results of the computations for the flipped $SU(5) \times U(1)_X$ models given in Table~\ref{tab:br}.

Observe in Table~\ref{MSSMSpectrum} containing the MSSM content charges for intersecting D-branes that there are no electrically neutral particles at the electroweak scale charged on $U(1)_B$, hence there are also no invisible decays in the D-brane model. Likewise, the only $Z'$ decay modes present then are $Z' \rightarrow jj$, $Z' \rightarrow t \bar{t}$, $Z' \rightarrow WW$, and $Z' \rightarrow Zh$. To calculate the decay widths and branching ratios, the $U(1)_{B} = U(1)_{LP}$ charges employed are $Q_{LP_{Q_L}} = 1/3$, $Q_{LP_{U_R}} = Q_{LP_{D_R}} = -1/3$, and $Q_{LP_{\mathcal{H}_u}} = 1$. As with the flipped $SU(5) \times U(1)_X$ models, the value of $g_{Z'}$ is constrained via the LHC constraints. We use tan$\beta = 25$ for D-branes also. The intersecting D-brane branching ratios are included in Table~\ref{tab:br}.

\begin{table}[htb] 
\centering
\begin{tabular}{c|c|c|c|c} \hline
$$&$Br(Z' \rightarrow jj)$ & $Br(Z' \rightarrow t \bar{t})$ & $Br(Z' \rightarrow W^+W^-)$ & $Br(Z' \rightarrow Zh)$\ \\ \hline \hline
${\rm flipped}~SU(5) \times U(1)_X$&$0.870$ & $0.078$ & $0.026$ & $0.026$\ \\ \hline
${\rm Intersecting~D-branes}$&$0.670$ & $0.130$ & $0.100$ & $0.100$\ \\ \hline
\end{tabular}
\caption{Computed branching ratios for $Z'$ boson decay for the four channels realized in leptophobic flipped $SU(5)\times U(1)_X$ and intersecting D-brane models. Given the absence of any electrically neutral particles at the electroweak scale in either model, there are no invisible decay modes.}
\label{tab:br}
\end{table}

The value of the $Z'$ gauge coupling $g_{Z'}$ is freely floated prior to application of the LHC constraints given in Eqs. (\ref{lhc_constraint1}) - (\ref{lhc_constraint3}). The individual cross-sections as a function of the $Z'$ gauge coupling are shown in Figure (\ref{fig:gzp_plot}) for both the flipped $SU(5) \times U(1)_X$ and intersecting D-brane models. The dashed and dotted lines in Figure (\ref{fig:gzp_plot}) represent all values of the gauge couplings, while the thick, solid sections are only those values of $g_{Z'}$ that satisfy the constraints of Eqs. (\ref{lhc_constraint1}) - (\ref{lhc_constraint3}). It is clear that implementation of the $Z' \rightarrow jj$ fitted $1 \sigma$ deviation very tightly constrains the value of the $U(1)_{LP}$ gauge coupling to $g_{Z'} \sim 0.25 - 0.45$ for the flipped models and $g_{Z'} \sim 0.50$ for D-branes. While these tight model constraints on $g_{Z'}$ are rather predictive, particularly for intersecting D-branes, it does leave little room for deviation in the event future enhancements of the constraints are necessary. Given the modest accumulation of only about 20 ${\rm fb}^{-1}$ of luminosity thus far at 8 TeV, and awaiting the forthcoming deluge of data extracted from the 13/14 TeV beam collision energies, we alternatively relax the $Z' \rightarrow jj$ constraint to only an upper limit on the cross-section of $\sigma(pp \rightarrow Z') \times Br (Z' \rightarrow jj) \lesssim 170~{\rm fb}$. This releases the lower values of $g_{Z'}$ as viable candidates, providing loose model constraints of $g_{Z'} \lesssim 0.49$ for flipped models and $g_{Z'} \lesssim 0.50$ for D-branes. The 13/14 TeV data allocation arriving in the year 2015 and beyond shall reduce the experimental measurement uncertainties and should ultimately merge the tight and loose model constraints presented here.




\begin{acknowledgments}

This research was supported in part by the Natural Science Foundation of China
under grant numbers 11135003, 11275246, and 11475238 (TL), and 
by the DOE grant DE-FG02-13ER42020 (DVN).

\end{acknowledgments}

\end{document}